\documentclass[12pt]{article}
\usepackage[english]{babel}
\usepackage{graphicx,epsfig}
\makeindex \textwidth 170mm \textheight 220mm \topmargin -5mm
\newcommand{\be}{\begin{equation}}
\newcommand{\ee}{\end{equation}}
\newcommand{\bea}{\begin{eqnarray}}
\newcommand{\eea}{\end{eqnarray}}

\def\bb{\bibitem}

\def\rp#1#2{{#1\over#2}}
\def\bar{\begin{eqnarray}}
\def\ear{\end{eqnarray}}
\def\eqi{\begin{equation}}
\def\eqf{\end{equation}}

\def\dert#1#2{\frac{d#1}{d#2}}

\def\rfr#1{eq.(\ref{#1})}
\def\rfrs#1#2{eqs.(\ref{#1})-(\ref{#2})}

\def\lb#1{\label{#1}}

\begin{document}
\begin{titlepage}
\begin{flushright}
\today\\
BARI-TH/00\\
\end{flushright}
\vspace{.5cm}
\begin{center}
{\LARGE Constraints to a Yukawa gravitational potential\\
from laser data to LAGEOS satellites} \vspace{1.0cm}
\quad\\
{Lorenzo Iorio$^{\dag}$\\
\vspace{0.5cm}
\quad\\
{\dag}Dipartimento di Fisica dell' Universit{\`{a}} di Bari, via
Amendola 173, 70126, Bari, Italy. E-mail: Lorenzo.Iorio@ba.infn.it}\\ \vspace{1.0cm}

{\bf Abstract\\}
\end{center}

{\noindent \footnotesize In this paper we investigate the
possibility of constraining the hypothesis of a fifth force at the
length scale of two Earth's radii by investigating the effects of
a Yukawa gravitational potential on the orbits of the
laser--ranged LAGEOS and LAGEOS II satellites. The existing constraints on the
Yukawa coupling $\alpha$, obtained by fitting the LAGEOS
orbit, are of the order of $| \alpha | < 10^{-5}-10^{-8}$ for
distances of the order of $10^9$ cm. Here we show that with a
suitable combination of LAGEOS and LAGEOS II data it should be possible
to constrain $\alpha$ at a level of
$4\times 10^{-12}$ or less. Various sources of systematic errors
are accounted for, as well. Their total impact amounts to $1\times
10^{-11}$ during an observational time span of 5 years. In the
near future, when the new data on the terrestrial gravitational
field will be available from the CHAMP and GRACE missions, these
limits will be further improved. The use of the proposed LARES
laser--ranged satellite would yield an experimental accuracy in
constraining $\alpha$ of the order of $1\times 10^{-12}$.}
\end{titlepage} \newpage \pagestyle{myheadings} \setcounter{page}{1}
\vspace{0.2cm} \baselineskip 14pt

\setcounter{footnote}{0}
\setlength{\baselineskip}{1.5\baselineskip}
\renewcommand{\theequation}{\mbox{$\arabic{equation}$}}

\noindent
\section{Introduction}
In this paper we investigate the constraints that can be posed on
the existence of a possible fifth force by the analysis of the
laser--ranged data to LAGEOS II and LAGEOS Earth satellites. We
will consider a potential energy including a Yukawa term of the
form [{\it Ohanian and Ruffini}, 1994; {\it Ciufolini and
Wheeler}, 1995; {\it Nordvedt}, 1998] \eqi U=U_0+U_{\rm Y
}=-\rp{GMm}{r}\left(1+\alpha
e^{-\rp{r}{\lambda}}\right),\lb{epot}\eqf where $G$ is the
Newtonian gravitational constant, $M$ is the mass of the central
body, $m$ is the mass of the orbiting test particle, $r$ is the
distance between the two bodies, $\alpha=\rp{Kk}{GMm}$ contains
the couplings $K$ and $k$ of the new force to the two
bodies\footnote{In general, $K$ and $k$ are not proportional to
$M$ and $m$.} and $\lambda$ is the finite range of the new force.

In checking the nature of the fifth force it is of the utmost
importance to perform experiments spanning the widest range of
length scales as possible [{\it Nordvedt}, 1998]: for experiments
at laboratory scale see [{\it Krause and Fischbach}, 2001] and
references therein. Using the orbits of LAGEOS and LAGEOS II
satellites implies
that we are testing the hypothesis of the fifth force at a length
scale of almost two Earth radii, i.e. 10$^4$ km: at this scale the
constraints on $\alpha$ are of the order of $ | \alpha |<
10^{-5}-10^{-8}$ (see Fig. 3.2 (a) of [{\it Ciufolini and
Wheeler}, 1995]) and are derived from a data analysis of LAGEOS.

Our analysis includes also an evaluation of the error budget in
order to account for various systematic errors induced by several
classical aliasing forces. We will show that it is possible to
improve sensibly the present limits by using suitably the data
from the existing LAGEOS and LAGEOS II satellites and the present
or near future knowledge of the terrestrial gravitational field
whose uncertainties represent, as we will see later, the main
sources of systematical errors.

The paper is organized as follows. In section 2 we derive the
effects of the Yukawa gravitational potential on the orbit of a
test body with the standard technique of the Gauss perturbative
equations for the rates of change of the Keplerian orbital
elements. In section 3 we apply the results obtained in section 2
to the Earth--LAGEOS system and discuss the constraints posed on
$\alpha$ by using a suitable combination of the residuals of the
perigee of LAGEOS II and the nodes of LAGEOS II and LAGEOS. The
effects of various sources of systematical errors are
investigated. The role of the proposed LARES satellite is
considered as well. Section 4 is devoted to the conclusions.
\section{The orbital effects of the Yukawa perturbation}
The acceleration felt by a test body orbiting the mass $M$ in the
potential energy given by \rfr{epot} is
\eqi\textbf{a}=\textbf{a}_{0}+\textbf{a}_{\rm
pert}=-\rp{GM}{r^2}\hat{r}-\alpha
GM\left(\rp{1}{r^2}-\rp{1}{\lambda^2}\right)\hat{r}.\lb{acc}\eqf
In obtaining \rfr{acc} it has been assumed that
$e^{-\rp{r}{\lambda}}\sim 1-\rp{r}{\lambda}$, as it should be the
case for an Earth orbiting satellite with $r\sim 10^7$ m. In
\rfr{acc} $\hat{r}$ is the unit vector pointing from the central
mass to the orbiting body.

The second term of the right--hand side of \rfr{acc} can be
considered a small, central perturbation of the Newtonian monopole
acceleration. It may be interesting to note that, since, in
general, $\alpha$ may vary from a body to another in the field of
the same mass\footnote{This happens if $k$ is not proportional to
$m$.}, \rfr{acc} implies that different bodies may be accelerated
differently in the field of $M$ [{\it Nordvedt}, 1998].

Let us work out explicitly the effects of the Yukawa perturbing
acceleration on the orbit of an artificial satellite. We will
adopt the standard approach based on the Gauss perturbative
equations and the projections of the disturbing acceleration $R,\
T,\ N$ onto the radial, along--track and cross--track mutually
orthogonal directions [{\it Milani et al.}, 1987], respectively.
The Gauss equations are \bar
\dert{a}{t} & = & \rp{2}{n\sqrt{1-e^2}}\left[Re\sin f+T\rp{p}{r}\right],\lb{smax}\\
\dert{e}{t} & = & \rp{\sqrt{1-e^2}}{na}\left[R\sin f+T\left(\cos f+\rp{1}{e}\left(1-\rp{r}{a}\right)\right)\right],\\
\dert{i}{t} & = & \rp{1}{na\sqrt{1-e^2}}N\rp{r}{a}\cos (\omega+f),\lb{in}\\
\dert{\Omega}{t} & = & \rp{1}{na\sin i\sqrt{1-e^2}}N\rp{r}{a}\sin (\omega+f),\lb{nod}\\
\dert{\omega}{t} & = & -\cos i\dert{\Omega}{t}+\rp{\sqrt{1-e^2}}{nae}\left[-R\cos f+T\left(1+\rp{r}{p}\right)\sin f\right],\lb{perigeo}\\
\dert{\mathcal{M}}{t} & = & n
-\rp{2}{na}R\rp{r}{a}-\sqrt{1-e^2}\left(\dert{\omega}{t}+\cos
i\dert{\Omega}{t}\right),\lb{manom} \ear
 where $a,\ e,\ i,\ \Omega,\ \omega$ and $\mathcal{M}$ are the
satellite's semimajor axis, eccentricity, inclination, longitude
of the ascending node, argument of perigee and mean anomaly,
respectively. Moreover, $p=a(1-e^2)$, $f$ is the true anomaly and
$n=\sqrt{GMa^{-3}}$ is the Keplerian mean motion. As can be
noticed from \rfr{acc}, the Yukawa disturbing acceleration has
only the in--plane, radial component $R$, so that it can be
straightforwardly inferred from \rfrs{in}{nod} that the
out--of--plane Keplerian orbital elements like the inclination $i$
and the longitude of the ascending node $\Omega$ are not affected
by it. By evaluating the Yukawa acceleration on the unperturbed
Keplerian ellipse, for which\eqi r=\rp{a(1-e^2)}{1+e\cos f} ,\eqf
inserting it in \rfrs{smax}{manom} and, then, averaging them over
an orbital revolution, the long--period rates of change of the
Keplerian orbital elements can be obtained. It turns out that, by
neglecting terms of order $\mathcal{O}(e^n)$, with $n\geq 2$, in
the satellite's eccentricity, only the perigee $\omega$ and the
mean anomaly $\mathcal{M}$ are affected by long--term Yukawa
perturbations. Indeed, from \rfr{acc} and \rfrs{perigeo}{manom} it
can be obtained \bar
\dert{\omega}{t} & = & \alpha\rp{n}{(1-e^2)^{\rp{3}{2}}},\lb{perig}\\
\dert{\mathcal{M}}{t} & = &
n+\alpha\rp{n}{(1-e^2)}-\alpha\rp{2GM(1-e^2)}{na\lambda^2}.\lb{man}\ear
\section{The constraint on ${\bf \alpha}$}
In order to constraint effectively the Yukawa coupling $\alpha$,
let us focus on \rfr{perig}. It tells us that the Yukawa
perturbation induces on the perigee of a near Earth satellite a
secular rate which, for LAGEOS II, is proportional to
$3.06658517\times 10^{12}$ milliarcseconds per year (mas/y) via
the coupling constant $\alpha$. What is the constraint posed on
$\alpha$ by the experimental accuracy with which it can be
possible to measure the perigee rate? In the case of this
Keplerian orbital element the observable quantity is $r=ea\omega$.
So, by assuming an experimental error of, say, $\delta r^{\rm II
}_{\rm exp}$ =1 cm over 1 year, for LAGEOS II, which has $e=0.014$
and $a=1.2163\times 10^{9}$ cm, we have $\delta\omega^{\rm
II}_{\rm exp}= 12$ mas\footnote{The other existing passive
geodetic laser--ranged satellites are unsuitable because their
eccentricities are smaller than LAGEOS II, except for Starlette
whose orbit, however, is known less accurately than that of LAGEOS
II for various reasons.}. This yields a relative accuracy on
$\alpha$ \eqi\left({\rp{\delta\alpha}{\alpha}}\right)^{\rm
exp}\sim 4\times 10^{-12}.\eqf It is interesting to note that the
same estimate for the proposed LARES satellite [{\it Ciufolini and
Matzner}, 1998], which is planned to have a larger eccentricity,
$e_{\rm LARES}=0.04$, would yield a relative accuracy of almost
$1\times 10^{-12}$.

However, this estimate does not include any systematic errors. As
it is well known from various proposed or performed tests of
General Relativity with LAGEOS satellites [{\it Ciufolini et al.},
1997; 1998; {\it Ciufolini}, 2000; {\it Iorio and Pavlis}, 2001;
{\it Iorio}, 2001; {\it Iorio et al.}, 2002; {\it Iorio}, 2002],
in such kind of measurements there are lots of competing classical
forces which may act as superimposed biases affecting sensibly the
precision of the measurements. Then, the evaluation of the
systematical errors induced by them is of the utmost importance.

The main source of systematic errors is represented by the
mismodelled precessions of the even zonal harmonics of the static
part of the geopotential. In particular, the first two even zonal
harmonics $J_2$ and $J_4$ are the most insidious. In order to
cancel their impact, as proposed in the PPN LAGEOS experiment
[{\it Iorio et al.}, 2002], the following
combination of orbital residuals could be used\footnote{Notice that, since
LAGEOS enters the combination of \rfr{combi} only with its node
$\Omega$, and since the Yukawa perturbation does not affect such
Keplerian orbital element, \rfr{combi} is insensitive to the
possible differential falling of LAGEOS and LAGEOS II in the
terrestrial gravitational field.} \eqi\delta\dot\omega^{\rm
II}+c_1\delta\dot\Omega^{\rm II}+c_1\delta\dot\Omega^{\rm
I}=\alpha x_{\rm Yuk},\lb{combi}\eqf where \bar c_1 & = & -0.86,\\
c_2 & = & -2.85,\\
x_{\rm Yuk} & = & 3.06658517\times 10^{12}\ {\rm mas/y}.\ear The
coefficients of \rfr{combi} depend on the orbital parameters of
LAGEOS and LAGEOS II and are obtained in order to cancel the
contributions of the first two even zonal harmonics of the
geopotential to the measurement of $\alpha$. It is intended that
the residuals would account for the Yukawa perturbation in the
sense that it would be viewed as an unmodelled feature not
included in the force models adopted in fitting the satellites'
orbits. According to the covariance matrix of the most recent
available Earth gravity model EGM96 [{\it Lemoine et al.}, 1998],
the systematic error induced by the uncancelled even zonal
harmonics of the geopotential amounts to\footnote{This estimate
has been obtained by considering the geopotential harmonics up to
degree $l=20$. This is well justified by the insensitivity of
LAGEOS satellites to the higher degree terms. Moreover, this fact
makes our estimate reliable because the higher degree terms of
geopotential in EGM96 are not particularly well determined.}
\eqi\left({\rp{\delta\alpha}{\alpha}}\right)^{\rm zonals}\sim
7\times 10^{-12}.\eqf Regarding the time--dependent part of the
Earth gravitational field, the estimates of [{\it Iorio et al.},
2002], adapted to this context, yield for an observational time
span of 5 years \eqi\left({\rp{\delta\alpha}{\alpha}}\right)^{\rm
harmonics}\sim 1\times 10^{-12}.\eqf The most relevant
non--gravitational perturbations are the direct solar radiation
pressure and the Earth's albedo. Their impact can be evaluated
from [{\it Lucchesi}, 2001] for 5 years as
\eqi\left({\rp{\delta\alpha}{\alpha}}\right)^{\rm non-grav}\sim
8\times 10^{-12}.\eqf The errors induced by the direct solar
radiation pressure and the Earth's albedo have been added
quadratically, as suggested in [{\it Lucchesi}, 2001]. Regarding
other subtle non--gravitational perturbations of thermal origin
acting on the orbits of the LAGEOS satellites, their effects on
the perigee of LAGEOS II are currently under accurate evaluation.

Then, a reliable estimate of the total systematic error induced by
various classical perturbations on the measurement of $\alpha$
through \rfr{combi} over a 5 years time span yields
\eqi\left({\rp{\delta\alpha}{\alpha}}\right)^{\rm total}\sim
1\times 10^{-11}.\lb{erto}\eqf In obtaining \rfr{erto} we have
summed in a root--sum--square fashion the gravitational and
non--gravitational errors assumed to be independent. Instead, the
gravitational error due to the static and time--dependent parts of
the Earth's gravitational field have been simply summed up in view
of a reciprocal correlation.
\section{Conclusions}
In this paper we have shown that by using a suitable combination
of the orbital residuals of the perigee of LAGEOS II and the nodes
of LAGEOS II and LAGEOS laser--ranged Earth satellites it would be
possible to constrain effectively the Yukawa coupling $\alpha$ of
a possible fifth force at a length scale of almost two Earth's
radii. The experimental sensitivity in measuring the perigee,
which would be affected by the Yukawa force, of LAGEOS II would
allow a precision on $\alpha$ of the order of almost $4\times
10^{-12}$ over a time span of 1 year. According to the present
knowledge of the terrestrial gravity field, the constraint posed
by the systematic errors is of the order of $1\times 10^{-11}$
over an observational time span of 5 years. These estimates should
greatly improve when the new and more accurate data on the Earth
gravity field from the CHAMP and GRACE missions will be available
in the near future. The use of the proposed LARES satellite, with
its larger eccentricity, would allow to improve the experimental
constraint to $1\times 10^{-12}$.

\end{document}